\documentstyle[eqsecnum,aps,twocolumn]{revtex}

\input epsf

\begin{document}
\draft
%\preprint{HIP-2000-xx/TH}
\title{Stochastic magnetohydrodynamic turbulence in space dimensions $d\ge 2$}
\author{M. Hnatich$^a$, J. Honkonen$^b$ and M. Jurcisin$^c$ }
\address{$^a$ Institute for Experimental Physics, SAS, Ko\v{s}ice, Slovakia
}
\address{$^b$ National Defence College and Department of Physics,
%P.O.Box 9
%(Siltavuorenpenger 20 C),\\FIN-00014
University of Helsinki, Helsinki, Finland
}
\address{$^c$ Joint Institute for Nuclear Research, Dubna, Russia}
%\author{J. Honkonen$^b$}
%\address{Theory Division, Department of Physics,
%P.O.Box 9
%(Siltavuorenpenger 20 C),\\FIN-00014
%University of Helsinki, Finland
%}
%\author{D. Horvath$^a$}
%\address{ Institute for Experimental Physics, SAS, Ko\v{s}ice, Slovakia
%}
%\author{R. Semancik$^c$}
%\address{P. J. \v{S}afarik University, Ko\v{s}ice, Slovakia}
\date{\today}
\maketitle

\begin{abstract}
Interplay of kinematic and magnetic forcing in a model of a conducting fluid
with randomly driven magnetohydrodynamic
equations has been studied in space dimensions $d\ge 2$
by means of the renormalization group.
A perturbative expansion scheme, parameters  of which are the
deviation of the spatial dimension from two and the deviation of the
exponent of the powerlike correlation function of random forcing
from its critical value, has been used in one-loop
approximation. Additional divergences have been taken into account
which arise at two dimensions and have been inconsistently treated in earlier
investigations of the model.
It is shown that in spite of the additional divergences the kinetic
fixed point associated with the Kolmogorov scaling regime remains stable
for all space dimensions $d\ge 2$ for rapidly enough falling off correlations
of the magnetic forcing. A scaling regime driven by thermal
fluctuations of the velocity field has
been identified and analyzed. The absence of a
scaling regime near two dimensions driven by the fluctuations of the magnetic field
has been confirmed.
A new renormalization scheme has been put forward and numerically investigated to interpolate
between the $\epsilon$ expansion and the double expansion.

\end{abstract}
\pacs{47.27.Gs,52.30.-q,11.10.Hi}
% \begin{center}
% {\bf Classification}
%
% \begin{verbatim}
%
% Isotropic turbulence, homogeneous turbulence   47.27.Gs.
% Turbulent diffusion                            47.27.Qb.
% High-Reynolds-numbers turbulence               47.27.Jv.
% Plasma flow, magnetohydrodynamics              52.30.-q.
% Renormalization group - field theory           11.10.G.
% Renormalization group evolution of parameters  11.10.Hi.
%
% \end{verbatim}
%
% \end{center}

\narrowtext

\section{Introduction}

During the past two decades asymptotic analysis of stochastic
transport equations [Navier-Stokes equation,
magnetohydrodynamic (MHD) equations, advection-diffusion equation and the
like] has attracted increasing attention.
Various forms of the renormalization group (RG) have proved to be
particularly useful in this investigation, and a great deal of work
has been carried out in the RG analysis of
the stochastic Navier-Stokes equation and the problem of
a passive scalar (turbulent diffusion
or heat conduction)~\cite{Frisch95,Adzhemyan99}.
Somewhat less effort has been devoted to the asymtotic analysis of
stochastic magnetohydrodynamics since the pioneering work of
Fournier et al~\cite{Fournier82} and Adzhemyan et
al~\cite{Adzhemyan85}. In particular, in these papers the existence of two
different anomalous scaling regimes (kinetic and magnetic) in three dimensions
was established corresponding to two non-trivial infrared-stable
fixed points of the renormalization group. It was also conjectured
that in two dimensions the magnetic scaling regime does not exist
due to instability of the magnetic fixed point. However, in both
papers there were flaws in the renormalization of the model in two
dimensions~\cite{Adzhemyan99,Honkonen96}.
Even more serious shortcomings are present in recent investigations of
MHD turbulence~\cite{Liang93,Kim99}, in which a
specifically two-dimensional setup has been applied with the use of the stream
function and magnetic potential. Therefore, results obtained for the
two-dimensional case in these papers cannot be considered completely conclusive.

In the present paper we have first carried out a field-theoretic
RG analysis of the stochastically forced
equations of magnetohydrodynamics with the proper account of
additional divergences which arise in two dimensions. This
gives rise to a two-parameter expansion of scaling exponents and
scaling functions~\cite{Honkonen96}, the parameters  of which are the
deviation of the spatial dimension from two and the deviation of the
exponent of the powerlike correlation function of random forcing
from its critical value. In this double
expansion the standard procedure of minimal subtractions was used
in the renormalization of the corresponding field-theoretic model.
We have carried out a one-loop
RG analysis of the large-scale asymptotic
behavior of the model and confirmed the basic conclusions of the
previous analyses~\cite{Fournier82,Adzhemyan85} that near two dimensions
a scaling regime driven by the velocity fluctuations may exist, but no
magnetically driven scaling regime can occur. We have also identified a
scaling regime driven by thermal fluctuations~\cite{Forster77} of the velocity
field.

Second, we have performed a renormalization of the model with a
different choice of finite renormalization in order to find at which
non-integer dimension the magnetic fixed point ceases to be
stable. This borderline dimension was found in Ref.~\cite{Fournier82}
with the use of the momentum-shell RG in a
setup valid in a fixed space dimension $d>2$. In the two-parameter expansion
with  the deviation of the
exponent of the powerlike correlation function of random forcing
from its critical value $\epsilon$ and
$2\delta=d-2$ as expansion parameters this effect cannot be
traced.
Therefore, we have carried out a RG analysis
according to the "principle of maximum divergences" in the sense
that we have included in the renormalization all graphs relevant in two dimensions,
and fixed the finite renormalization in a way which reproduces the
results of a momentum-shell renormalization of Wilson~\cite{Wilson74} at
one-loop order. This procedure gives rise to
RG functions such that in the limit of small
$\delta$, $\epsilon$ they reproduce the results of the two-parameter
expansion,
and in the limit of small $\epsilon$ (but finite $\delta$) they
yield the results of the usual $\epsilon$
expansion~\cite{Fournier82,Adzhemyan85}.

We have also investigated
the long-range asymptotic behavior of the model in the framework of
the latter scheme without any small parameter and found, in particular, that in this
case thermal
fluctuations make the value of  the borderline
dimension of the magnetic scaling regime significantly lower
($d_c=2.46$) than in the $\epsilon$ expansion~\cite{Fournier82})
($d_c=2.85$).

This paper is organized as follows:
Section II starts from the functional
formulation of the solution of stochastic MHD. This is convenient
 for the analysis based
 on the standard field-theoretic RG approach, the details of which are
 described in Sec. III.
The Kolmogorov constant for MHD is calculated in Sec. IV at the
leading order of
the two-parameter expansion. In Sec. V a RG analysis of the
model with
maximum divergences is carried out in arbitrary dimension, and the
observed strong effect of thermal fluctuations is discussed.
In Sec. VI the conclusions are presented.

\section{Field theory for stochastic magnetohydrodynamics}

We consider the model of stochastically forced conducting fluid
 described by the system of magnetohydrodynamic equations
 for the fluctuating velocity field ${\bf v} (t,{\bf x})\equiv{{\bf v}}(x)$
 of an incompressible conducting fluid
 and the magnetic induction ${\bf B}=\sqrt{\rho\mu}{\bf b}$
 ($\rho$ is the density and $\mu$ the permeability of the fluid) in the form
 \cite{Fournier82,Adzhemyan85}
 \begin{eqnarray}
 &&  \partial_t {\bf v} +
P \left[({\bf v}\cdot {\nabla} ) {{\bf v}}-
({{\bf b}}\cdot{\nabla}){{\bf b}}\right] -
 \nu_0 \nabla^2 {{\bf v}}
  ={{\bf f}}^v\,,
  \label{NSR}\\
 && \partial_t {\bf b} +
 ({\bf v}\cdot {\nabla}){\bf b}
 - ({\bf b}\cdot {\nabla}){\bf v}
 -\nu_0 u_0\nabla^2 {{\bf b}}  = {\bf f}^b\,, \label{magnet}
 \end{eqnarray}
together with the incompressibility conditions
 \begin{equation}
 \label{NSR1}
 {\nabla}\cdot{{\bf v}}   = 0,\qquad
 {\nabla}\cdot{{\bf f}}^v = 0,\qquad
  {\nabla}\cdot{{\bf f}}^b = 0.
 \nonumber
 \end{equation}
 In (\ref{NSR}) $P$ is the transverse projection operator, $\nu_0$
 the (unrenormalized) kinematic viscosity and $1/u_0$ the
 unrenormalized magnetic Prandtl number. In statistical applications of the
 field-theoretic RG the unrenormalized (bare) paramaters are the
 physical ones.

 The statistics of ${{\bf v}}$ and ${{\bf b}}$
 are completely determined by the non-linear equations
 (\ref{NSR}), (\ref{magnet}),
 (\ref{NSR1}),
 and the probability distribution
of the external large-scale random forces
 ${\bf f}^v$, ${\bf f}^b$.
It is customary~\cite{Fournier82,Adzhemyan85} to consider random
forces ${{\bf f}}^v$ and ${{\bf f}}^b$ having a zero-mean Gaussian
distribution with correlation functions of the form
\begin{equation}
 D_{mn}(x) = \delta(t)\!
 \int \frac{\mbox{d}^d {{\bf k}}}{(2\pi)^d}\,
 e^{i{{\bf k}}\cdot{{\bf x}}}D(k)\left(\delta_{mn}-{k_mk_n\over
 k^2}\right)\,,
 \label{Cor0}
 \end{equation}
 in which the
 time correlations of the fields
 have the character of white noise, while the spatial
 correlations are controlled by the
 scalar function $D(k)$. Transversality of the matrix (\ref{Cor0})
 is a consequence of the equations
 $\nabla\cdot{{\bf v}}=\nabla\cdot{{\bf b}}=0$.

 To analyze renormalization near two dimensions
 we consider the model  (\ref{NSR}), (\ref{magnet}), (\ref{NSR1})
 supplemented by the forcing statistics
 \begin{eqnarray}
 \label{corel}
 \langle \,f_m^v(x_1) f_n^v(x_2)
 \,\rangle&=&
 u_0\,\nu_0^3\,D_{mn}
\left( x_1-x_2;\{ 1, g_{v10}, g_{v20}
\}\right)\,,\nonumber\\
 \langle\, f_m^b(x_1) f_n^b(x_2)
 \,\rangle&=&
 u_0^2\,\nu_0^3\,D_{mn}
\left( x_1-x_2;\{ a, g_{b10}, g_{b20} \}\right)\,, \nonumber\\
 \langle\, f_m^v(x_1) f_n^b(x_2)\,\rangle&=&0\,,
 \end{eqnarray}
 where
 \begin{eqnarray}
 \label{silaN}
 D_{mn}\left( x; \{ A,B,C \} \right) =
\delta( t )\nonumber\\
\times\int
\frac{{\mbox d}^d {{\bf k}}}{(2\pi)^d}
\,\,
P_{mn}({{\bf k}})
e^{i{{\bf k}}\cdot{{\bf x}}}
\left[
B\, k^{ 2- 2\delta-2\,A\, \epsilon}
+
C\,k^2\,\right]\,.
\end{eqnarray}
  All the dimensional constants
 $g_{v10}$, $g_{b10}$, $g_{v20}$ and $g_{b20},$ in~(\ref{corel})
 control  the amount of randomly
 injected energy.
 The choice of the values of the parameters $\epsilon$ and $\delta$
 determines the powerlike falloff of the long-range forcing correlations
 and the space dimension of the system under consideration.

 We choose uncorrelated kinematic and magnetic driving [$ \langle\, {\bf f}^v
 {\bf f}^b\,\rangle=0$], because we are considering arbitrary
 space dimension $d\ge 2$ and it is not possible to define a
 nonvanishing correlation function of a vector field and a pseudovector field in
 this case. This can be done separately for integer dimensions of space,
 but, contrary to claims of some
 authors~\cite{Fournier82,Camargo92}, is no obstacle for
 application of the RG~\cite{Adzhemyan85}.

The structure of the  matrix $D_{mn}$ in (\ref{silaN}) reflects
 a more detailed intrinsic
 statistical definition of forcing, whose consequences
 are deeply discussed in Refs.~\cite{Adzhemyan99,Horvath95}.
 Technically, it is necessary to accompany the long-range correlations
 [corresponding to the term $k^{ 2- 2\delta-2\,A\, \epsilon}$ in
 (\ref{silaN})]
 by local correlations [described by the analytic in $k^2$ term in the correlation
 function (\ref{silaN})] in order to construct a consistent
 renormalization procedure for the corresponding field-theoretic
 model~\cite{Honkonen96,Honkonen98,Hnatich99}. This feature has been
 overlooked in the previous analyses of the
 problem~\cite{Fournier82,Adzhemyan85}.
 The prefactors $ u_0\,\nu_0^3$ and
 $ u_0^2\,\nu_0^3$ in (\ref{corel}) have been extracted for the
 convenience of calculations.

 The definition (\ref{silaN}) includes two principal -- low-
 and high-wave-number scale -- kinetic forcings separated
 by transition region in the vicinity
 of the characteristic wave number of order
 $O([B/C]^{1/(2\delta+2A\epsilon)}).$
 The forcing contribution with local
 correlations gives a
 phenomenological description of small-scale thermal fluctuations
 of the magnetic induction and the velocity field~\cite{Forster77}.

The long-range parts of the translational invariant correlation
functions (\ref{silaN}) become scale invariant at the values $\epsilon=2$, $a=1$.
 For the exponent $\epsilon$ the value $\epsilon=2$ is physically most
 reasonable, since it represents the assumption
 that random forces in the Navier-Stokes equation~(\ref{NSR})
 act at very large scales,
 which substitutes for the effect of boundary conditions.

We are working in an arbitrary dimension, but the renormalization will be carried out
within the two-dimensional model. In two-dimensional magnetohydrodynamic turbulence, in contrast to fluid
turbulence, there are direct energy cascades in both two and
three dimensions. Therefore it is natural to expect that the
scaling behavior is rather similar in both cases, and
we apply the same forcing spectrum in all space dimensions $d\ge 2$.

We use the correlation functions
 \begin{equation}
  \label{Cor2}
 \langle
 \,
 v_{j_1} (x_1)
 v_{j_2} (x_2)
 v_{j_3} (x_3)
 \cdots v_{j_N}(x_N)
 \,
 \rangle\,,
  \end{equation}
where
$ 1\leq j_r\leq d$,
$r=1,2,\cdots N$
as measurable quantities for the description of  turbulence
statistics. We have applied the RG method to the calculation of
 asymptotic properties of the correlation functions in
 the way initiated in Ref.~\cite{Vasilyev83}.
This approach is based on
a formal mapping from the stochastic
model (\ref{Cor0}) to a quantum-field model~\cite{DeDominicis76,Janssen76} with
a De-Dominicis-Janssen action $S\{{{\bf v}},{{\bf v}}',{{\bf b}},{{\bf
 b}}'\}$, which is a functional of the physical fields ${{\bf
 v}}$, ${{\bf b}}$  and independent solenoidal auxiliary fields
 ${{\bf v}'}$, ${{\bf b}'}$.
Thus, the correlation functions (\ref{Cor2})
can be expressed as functional averages with the "weight
functional"
$
 {\cal W}=\exp S
$.
The system of the stochastic MHD equations
 (\ref{NSR}), (\ref{magnet}), (\ref{NSR1}), (\ref{corel}), and (\ref{silaN})
gives rise to the following De-Dominicis-Janssen action:
\begin{eqnarray}
 \label{ACTION}
&S&=\frac{1}{2}
\int\!\mbox{d} x_1 \int\! \mbox{d} x_2\,\nonumber\\
&\times&
\Bigl[u_0\,\nu_0^3\,v'_m(x_1)
D_{mn}\left( x_1- x_2; \{ 1 , g_{v10}, g_{v20}\} \right)
v'_n(x_2)\nonumber\\
&+&
u_0^2\,{\nu_0}^3\,
b'_m(x_1)\,
D_{mn}\left( x_1 - x_2; \{ a, g_{b10}, g_{b20}\} \right)\,
b'_n(x_2)\,
\Bigr]\nonumber\\
&+&
\int\! \mbox{d} x
\,\Bigl\{{\bf v}'\cdot\left[  -\partial_t {\bf v} -
({\bf v}\cdot {\nabla} ) {{\bf v}} +
\nu_0 \nabla^2 {\bf v}+
({{\bf b}}\cdot{\nabla}) {{\bf b}}
\right]\nonumber\\&+&
{{\bf b}}'\cdot
\left[
 - \partial_t {{\bf b}}
 + u_0\,\nu_0 \nabla^2 {{\bf b}}
 + ({{\bf b}}\cdot{\nabla}) {{\bf v}}
 - ({{\bf v}}\cdot{\nabla}) {{\bf b}}
 \right]\Bigr\}\,.
 \end{eqnarray}
 The dimensional constants
 $g_{v10}$, $g_{b10}$, $g_{v20}$, and $g_{b20}$, which
 control  the amount of randomly
 injected energy through (\ref{corel}),
 play the role of expansion parameters of the perturbation theory.

\section{Two-parameter expansion of the model}

The action~(\ref{ACTION}) gives rise to
four three-point interaction vertices defined by the standard rules~\cite{Justin89}, and
the following set of propagators
 \begin{eqnarray}
  \label{PropY}
 \Delta^{v v'}_{mn}({{\bf k}},t)&=&
 \Delta^{v'v}_{mn}(-{{\bf k}},-t)
 = \theta(t)\,P_{mn}({{\bf k}})\,e^{-\nu_0\,k^2 t}\,,
  \nonumber\\
 \Delta^{bb'}_{mn}({{\bf k}},t)&=&
 \Delta^{b'b}_{mn}(-{{\bf k}},-t)
 = \theta(t)\,P_{mn}({{\bf k}})\,
e^{- u_0 \nu_0 k^2 t }\,,
 \nonumber\\
 \Delta^{vv}_{mn}({{\bf k}},t)
 &=&\frac{1}{2}
 \,u_0\,\nu_0^2\,P_{mn}({{\bf k}})\,
 e^{ -\nu_0 k^2 | t | }\nonumber\\
 &\times& \left(\,
 g_{v10}\,k^{-2\epsilon-2\delta}+g_{v20}
 \,\right)\,,
\\
 \Delta^{bb}_{mn}({{\bf k}},t)&=&
 \frac{1}{2}
 \,u_0\,\nu_0^2\,P_{mn}({{\bf k}})\,
 e^{ -u_0 \nu_0 k^2 | t | }\nonumber\\
 &\times& \left(\,
 g_{b10}\,k^{-2\,a \epsilon- 2\delta} + g_{b20}
 \,\right)
 \nonumber
 \end{eqnarray}
in the time-wave-number representation.
With due account of Galilei invariance of the action~(\ref{ACTION}), and
careful analysis of the structure of the perturbation expansion it can be shown~\cite{Adzhemyan85}
that for any fixed space dimension $d>2$ only
three one-particle irreducible (1PI) Green functions
$\Gamma^{vv'}$, $\Gamma^{bb'}$ and $ \Gamma^{v'bb}$
with superficial UV divergences are generated by the
action. They give rise to counter terms of the form
already present in the action, which thus is multiplicatively renormalizable by
power counting for space dimensions $d>2$.

We would like to emphasize that the structure of renormalization
should always be analyzed separately and it is not at all obvious
that the nonlinear terms are not renormalized in the solution of
the stochastic MHD equations. In fact, direct calculation shows
that the Lorentz-force term is renormalized. There seems to be a
certain amount of confusion about this point in the recent
literature. For instance, in Refs.~\cite{Liang93,Camargo92} the
authors erroneously neglect renormalization of nonlinear terms
as high-order effect. The approach adopted in Ref.~\cite{Liang93}
for two-dimensional MHD turbulence
was quite recently criticized by Kim and Young~\cite{Kim99}, who,
alas, in their field-theoretic treatment of the same problem ignore
renormalization of the Lorentz force without any justification.
They also neglect renormalization of the forcing correlations
by effectively considering renormalization of the model at
$d>2$, which does not seem to be appropriate in a setup in which
the strictly two-dimensional quantities, the stream function and magnetic
potential, are used for the description of the problem.

The analysis of the autocorrelation functions of the velocity
field and magnetic induction is essential near two dimensions,
since
in two dimensions additional divergences in the graphs of the 1PI
Green functions $\Gamma^{v'v'}$ and  $\Gamma^{b'b'}$ occur.
The point here
is~\cite{Honkonen96,Honkonen98} that
the nonlocal term of the action (and the similar one with the
auxiliary field ${\bf b}'$)
\begin{eqnarray*}
\int\!\mbox{d}t  \int\! \mbox{d}^{d} {\bf x}_1
 \int\! \mbox{d}^{d} {\bf x}_2\,
 {\bf v}'({\bf x}_1, t )\cdot
 {\bf v}'({\bf x}_2, t )\,\\
\times \int\! \frac{d^d \bf k}{(2\pi)^d}\,
 k^{2 - 2\delta-2\epsilon}\,
 e^{ i {\bf k} \cdot ( {\bf x}_1 - {\bf x}_2 )}
\end{eqnarray*}
brought about by the force correlation functions~(\ref{corel})
is not renormalized since
the divergences produced by the loop integrals of the graphs are always local
in space and time~\cite{Justin89}.
The simplest way to include the corresponding local counter terms
${\bf v}'\nabla^2 {\bf v}'$ and ${\bf b}'\nabla^2 {\bf b}'$
in the renormalization is to add corresponding
{\em local} terms  to the force
correlation function at the outset in order to keep the model multiplicatively
renormalizable, which is convenient technically. This is why
we have used the force correlation functions~(\ref{corel}) and
(\ref{silaN}) with both long-range and short-range correlations.
As a result, the action~(\ref{ACTION}) is multiplicatively
renormalizable and allows for a standard RG asymptotic analysis~\cite{Justin89}.

In the momentum-shell analysis of Fournier et
al.~\cite{Fournier82} these divergences were taken into account
only in the special case, when the force correlation function (\ref{silaN})
is local, i.e. $\propto k^2$ (formally this was fixed by the
condition $2\delta+2A\epsilon=0$, which excludes
$A\epsilon$ from the parameters of the model).
In the field-theoretic treatment of Adzhemyan et
al.~\cite{Adzhemyan85} the contribution of the
additional divergences was prescribed to a renormalization of
the non-analytic term $\propto  k^{2 - 2\delta-2\epsilon}$,
although only analytic in $k^2$ terms are produced in the course of
renormalization.

The model is regularized using a combination of analytic and
dimensional regularization with the parameters $\epsilon$ and
$2\delta=d-2$. As a consequence, the UV divergences appear as
poles in the following linear combinations of the regularizing
parameters: $\epsilon$, $\delta$, $2\epsilon+\delta$, and
$(a+1)\epsilon+\delta$.
The UV divergences  can be
removed by adding suitable counterterms to the basic action
$S_B$ obtained from the unrenormalized one (\ref{ACTION}) by
the substitution of the renormalized
parameters for the bare ones:
$g_{v10}  \rightarrow \mu^{2\epsilon}  g_{v1}$,
$g_{v20}  \rightarrow \mu^{-2 \delta} g_{v2}$,
$g_{b10}  \rightarrow \mu^{2 a \epsilon} g_{b1}$,
$g_{b20}  \rightarrow \mu^{-2 \delta} g_{b2}$,
$\nu_0 \rightarrow \nu$,
$u_0 \rightarrow u$,
where $\mu$ is a scale setting parameter having the same canonical
dimension as the wave number.

To construct an analog of the usual
$\epsilon$ expansion with $\epsilon$ and $\delta$ as small
parameters of the same order of magnitude, it is convenient to use
the minimal-subtraction (MS) scheme for the renormalization. In this
approach only singular parts of the Laurent series of the
superficially divergent 1PI Green functions are included in the
renormalization constants, which give rise to the counter terms
added to the basic action to make the Green functions of the
resulting renormalized model UV finite. The counter terms for
the basic action
corresponding to the unrenormalized action~(\ref{ACTION}) are
\begin{eqnarray}
\label{kontra}
\Delta S&=&\int\,\mbox{d} x\,
\big[ \nu\, \left(Z_1-1\right) {{\bf v}}'\nabla^2\, {{\bf v}}+
     u \nu\,\left(Z_2-1\right) {{\bf b}}'\nabla^2\, {{\bf b}}
\nonumber\\
&+&\frac{1}{2}\,(1-Z_4)
u {\nu^3} g_{v2}\,{\mu}^{-2\delta}\,
{{\bf v}}'{\nabla^2} {{\bf v}}'\nonumber\\
&+&\frac{1}{2}\,
(1- Z_5 )\,u^2 {\nu^3} g_{b2}\,{\mu}^{-2\delta}\,
{{\bf b}}'{\nabla^2} {{\bf b}}'
\nonumber\\
&+&(Z_3-1)\,
{{\bf v}}'({{\bf b}}\cdot{\nabla})\, {{\bf b}}\, \big]\,,
\end{eqnarray}
where the renormalization constants $Z_1$, $Z_2,$
$Z_4$, $Z_5$  renor\-mali\-zing the unrenormalized (bare)
para\-me\-ters
$e_0=\{g_{v10},g_{v20},g_{b10},g_{b20},u_0,\nu_0\}$ and the constant  $Z_3$ renormalizing
the fields ${\bf b}$, and ${{\bf b}}'$, are
chosen to cancel the UV divergences appearing in the Green functions constructed
using the basic action.
Due to the Galilean invariance of the action
the fields ${{\bf v}}'$, and ${{\bf v}}$ are not renormalized.

 In a multiplicatively renormalizable model, such as~ (\ref{ACTION}),
 the counter terms (\ref{kontra})
 can be chosen in a form containing a finite number of
 terms of the same algebraic structure
 as the terms of the original action (\ref{ACTION}).
 Thus, all UV divergences of the graphs of the perturbation expansion
 may be eliminated
 by a redefinition of the parameters of the original model.

 Renormalized Green functions are expressed
 in terms of the renormalized parameters
 \begin{eqnarray}
  \label{Zkaka1}
 g_{v1}&=&g_{v10}\,{\mu}^{-2\epsilon}\, Z_1^2 Z_2,\,\,\,
 g_{v2}=g_{v20}\,{\mu}^{2\delta}\, Z_1^2 Z_2 Z_4^{-1},
 \nonumber\\
  \nu&=&\nu_0\, Z_{1}^{-1},
 \quad u=u_0\, Z_2^{-1} Z_1\,,\\
 g_{b1}&=&g_{b10}\,{\mu}^{-2\,a\epsilon}\, Z_1 Z_2^{2} Z_3^{-1},\,\,\,
 g_{b2}=g_{b20}\,{\mu}^{2\delta}\, Z_1 Z_2^{2} Z_3^{-1} Z_5^{-1},\nonumber
 \end{eqnarray}
which are the parameters of the renormalized action
$S_R=S_B+ \Delta S$
connected with the unrenormalized action (\ref{ACTION})
by the relation of multiplicative renormalization
 \[
 S_{R}\left\{{\bf v},{\bf b} ,
 {\bf v}',{{\bf b}}',e\right\} =
 S\left\{{\bf v},
 {{\bf b}} Z_3^{\frac{1}{2}},{{\bf v}}',
 {{\bf b}}'Z_3^{-\frac{1}{2}},e_0\right\}\,,
 \]
 where $e$ is a shorthand for all the renormalized parameters
$ g_{v1}, g_{v2}, g_{b1}, g_{b2}, u$, and  $\nu $. Calculation of the correlation
and response functions of the velocity and magnetic fields with the use
of the renormalized action
yields renormalized Green functions without  UV divergences.

Independence of the unrenormalized Green functions of the
scale-setting parameter $\mu$ may be expressed in the form of
differential RG equations for the renormalized 1PI Green functions.
To keep notation simple, we quote these equations only for the
renormalized pair
correlation functions of the velocity field and the magnetic
induction. We define the Fourier transforms as
 \begin{eqnarray*}
 W^{vv}_{R\ mn}(t_1-t_2,{\bf k};g)=\int\,
 \frac{\mbox{d}^d{\bf x}_1}{(2\pi)^{d}}\,\\
\times \langle\,v_m({\bf x}_1,t_1)\,v_n({\bf x}_2,t_2)\,\rangle
 \,e^{ i {\bf k} \cdot ({\bf x}_1-{\bf x}_2)}\,,
 \\
 W^{bb}_{R\ mn}(t_1-t_2,{\bf k};g)=\int\,
 \frac{\mbox{d}^d{\bf x}_1}{(2\pi)^{d}}\,\\
 \times\langle\,b_m({\bf x}_1,t_1)\,b_n({\bf x}_2,t_2)\,\rangle
 \,e^{ i {\bf k} \cdot ({\bf x}_1-{\bf x}_2)}\,.
 \end{eqnarray*}
The basic RG equations for the correlation functions are
\begin{eqnarray}
\label{CS}
 \Biggl[\,{\mu}
 {\partial\over\partial \mu} +\beta_g{\partial\over\partial g}
-\gamma_1\nu{\partial\over\partial \nu}\,\Biggr]\,
W^{vv}_{R\ mn}=0\,,\nonumber\\
 \Biggl[\,{\mu}
 {\partial\over\partial \mu} +\beta_g{\partial\over\partial g}
-\gamma_1\nu{\partial\over\partial \nu}+\gamma_3\,\Biggr]\,
W^{bb}_{R\ mn}=0\,,
\end{eqnarray}
where $\beta_g\partial_g $ is a shorthand for
\[
 \beta_g{\partial\over\partial g}\! =\!
 {\beta}_{gv1}{\partial\over\partial g_{v1}} +
 {\beta}_{gv2}{\partial\over\partial g_ {v2}} +
 {\beta}_{gb1}{\partial\over\partial g_{b1}} +
 {\beta}_{gb2}{\partial\over\partial g_{b2}} +
 {\beta}_u{\partial\over\partial u}.
\]
The coefficient functions of
Eqs.~(\ref{CS}) $\beta_g$ and $\gamma_1$ are expressed in terms of
logarithmic derivatives
of the renormalization constants. We use the definitions
 \begin{equation}
 \gamma_i=\mu \frac{\partial \ln Z_i}{\partial \mu}\Biggl|_{0}\,,\,\,
\,\,\,
 \beta_g=\mu \frac{\partial g}{\partial\mu }\Biggl|_{0}\,,
 \,\,
 \label{defbeta}
 \end{equation}
where $ g=\{g_{v1},g_{v2},g_{b1},g_{b2},u\}$,
and the subscript "0" refers to partial derivatives
taken at fixed values of the bare parameters $e_0$.
It should be noted that here the functions $\beta_g$ and $\gamma_1$ are
functions of the parameters $g$ only.

Expressing the correlation functions through dimensionless scalar
functions $R_v$ and $R_b$ as
\begin{eqnarray*}
W_{R\ mn}^{vv}(t,{\bf k};g)=\nu^2k^{-2\delta}P_{mn}({\bf k})R_v(\tau,s;g)\,,\\
W_{R\ mn}^{bb}(t,{\bf k};g)=\nu^2k^{-2\delta}P_{mn}({\bf k})R_b(\tau,s;g)\,,
\end{eqnarray*}
where
$s=k/\mu$, $s\in [ 0,1 ]$ is the dimensionless wave number, and $\tau=t\nu k^2$
the dimensionless time, and
solving Eqs. (\ref{CS}) by the method of characteristics, we obtain
the correlation functions in the form
\begin{eqnarray}
\label{Rsolution}
W_{R\ mn}^{vv}(t,{\bf k};g) &=&
P_{mn}({\bf k})\, \overline{\nu}^2k^{-2\delta}
R_v\left(tk^2\overline{\nu},1; {\overline g}\right)\,,\nonumber\\
W_{R\ mn}^{bb}(t,{\bf k};g) &=&
e^{\int_1^s\!{\rm d}x\,\gamma_{3}({\overline
g}(x))/x}\\
&\times& P_{mn}({\bf k})\,\overline{\nu}^2k^{-2\delta}
R_b\left(tk^2\overline{\nu},1; {\overline g}\right)\,,\nonumber
\end{eqnarray}
where  ${\overline g}$ is the solution of the Gell-Mann-Low equations:
\begin{equation}
\label{Gell}
\frac{d\overline{g}(s)}{d\ln s}= \beta_g
\left[ \overline{g}(s)\right]\,,
\end{equation}
and $\overline{\nu}$ is the running coefficient of viscosity
\[
\overline{\nu}=\nu e^{-\int_1^s\!{\rm d}x\,\gamma_{1}({\overline
g}(x))/x}\,.
\]
The scale-invariant asymptotic behaviour of
the correlation functions stems from the
existence of a stable fixed point of the RG
transformation $\beta_g=0$ determined by the
Gell-Mann-Low equations (\ref{Gell}).

The definitions  (\ref{defbeta}) and the relations  (\ref{Zkaka1})
yield $\beta$ functions of the form
 \begin{eqnarray}
  \label{beee1}
 \beta_{gv1}&=&g_{v1}\,(-2\epsilon+2\gamma_1+\gamma_2)\,,\nonumber\\
 \beta_{gv2}&=&g_{v2}\,(2\delta+2\gamma_1+\gamma_2-\gamma_4)\,,
 \nonumber\\
 \beta_{gb1}&=&g_{b1}\,(-2\,a\epsilon+\gamma_1+2\gamma_2-\gamma_3)\,,\\
 \beta_{gb2}&=&g_{b2}\,(2\delta+\gamma_1+2\gamma_2-\gamma_3-\gamma_5)\,,\nonumber\\
 \beta_u&=& u\,(\gamma_1-\gamma_2)\,.\nonumber
 \end{eqnarray}
At one-loop accuracy the $\gamma$ functions are
\begin{eqnarray}
\label{GGGa1}
\gamma_1&=&\frac{1}{32\pi}[u\,(g_{v1}+g_{v2})+g_{b1}+g_{b2}]\,,\nonumber\\
\gamma_2&=&\frac{1}{8\pi}\frac{g_{v1}+g_{v2}-g_{b1}-g_{b2}}{u+1}\,,
\nonumber\\
\gamma_3&=&\frac{1}{16\pi}(g_{b1}+g_{b2}-g_{v1}-g_{v2})\,,\\
\gamma_4&=&\frac{1}{32\pi}\frac{u\,(g_{v1}+g_{v2})^2+(g_{b1}+g_{b2})^2}{g_{v2}}\,.\nonumber
\end{eqnarray}
There are no UV divergences
 in the 1PI Green function $\Gamma^{b'b'}$
 in the one-loop approximation, therefore
 \begin{equation}
 \gamma_5=0, \qquad Z_5 = 1,
 \label{G5}
 \end{equation}
 which is a specific property of the two-dimensional MHD.

Large-scale
asymptotic behavior is governed by infrared stable fixed points of
Eqs. (\ref{Gell}), determined by
the system of equations $\beta_{g}(g^{\ast})=0, $ and the conditions
$\overline{g}\to g^{\ast}$, when $s\rightarrow  0$.
For ${\overline g}(s)$ close to $g^{\ast}$ we obtain a system
 of linearized equations
 \[
 \left(
 I\,s\,\frac{\mbox{d}}{\mbox{d} s}
 - \Omega\,
 \right) \,({\overline g} - g^{\ast})\,= 0,
 \]
 where $I$ is the $5\times 5$ unit matrix and the matrix
 $\Omega= (\partial \beta_g/\partial g)|_{g^{\ast}}$. Solutions of this system
 behave like ${\overline g}=g^{\ast}+{\cal O}( s^{{\lambda}_j}) $,
 when $s\rightarrow 0$.
 The exponents $\lambda_j$, $j=1,2,3,4,5$
 are the eigenvalues of the matrix $\Omega$.
 In the vicinity of the fixed
point all the trajectories $g(s)$ approach
the fixed point, if the matrix $\Omega$
 is positive definite
[i.e. $\mbox{Re}\,(\lambda_j)> 0$].

Apart from the Gaussian fixed point $g_{v1}^*=g_{v2}^*=g_{b1}^*=g_{b2}^*=0$
with no fluctuation effects on the large-scale asymptotics,
which is IR stable for $\delta>0$, $\epsilon<0$, $a>0$,
there are two nontrivial  IR stable fixed
points of the RG with nonnegative $g_{v1}^*$,  $g_{v2}^*$, $g_{b1}^*$,
$g_{b2}^*$, and $u^*$.

The thermal fixed point is generated by
short-range correlations of the random
force with
 \begin{eqnarray}
 \label{thermal}
  g_{v1}^{\ast}   = 0,\quad
  g_{v2}^{\ast}    = -4\pi (1 + \sqrt{17} ){\delta}, \nonumber\\
  g_{b1}^{\ast}    = 0, \quad
  g_{b2}^{\ast}    = 0,
 \end{eqnarray}
and the inverse magnetic Prandtl number
\begin{equation}
\label{Pr}
 u^{\ast}=\frac{\sqrt{17}-1}{2}\simeq 1.562\,.
\end{equation}
Physically, the asymptotic behavior described by this fixed point
is brought about by thermal
fluctuations of the velocity field~\cite{Forster77}.
The region of stability of the thermal fixed point (\ref{thermal}), (\ref{Pr}) is
$2\epsilon+3\delta<0$, $\delta<0$ in the $\delta$, $\epsilon$
plane. For the magnetic forcing-decay parameter
$a$ the stability region is determined by the inequality
$8a\epsilon+(13+\sqrt{17})\delta<0$.

The {\em kinetic} fixed point~\cite{Fournier82} generated by the
forced fluctuations of the velocity field
is given by the universal inverse magnetic Prandtl number (\ref{Pr}),
the parameters
\begin{eqnarray}
g_{v1}^{\ast}=
\frac{128\,\pi}{9(\sqrt{17}-1)}\,\frac{\epsilon\,(2\epsilon+3\delta)}
{\epsilon+\delta}
,\nonumber\\
g_{v2}^{\ast}=\frac{128\pi}{9(\sqrt{17}-1)}\,
\frac{\epsilon^2}{\delta+\epsilon}\,,
\label{fix11}
\end{eqnarray}
and zero couplings of the magnetic forcing
 \begin{displaymath}
 g_{b1}^{\ast}=g_{b2}^{\ast}=0\,,
 \end{displaymath}
and it may be associated with
turbulent advection of the magnetic field.
The values of $g_{v1}^{\ast}$ and $g_{v2}^{\ast}$ in
(\ref{fix11})
correspond to those find previously in Ref.~\cite{Honkonen96}.
The region of stability
of the kinetic fixed point in the $\delta$, $\epsilon$ plane
is $\epsilon>0$, $2\epsilon+3\delta >0$.
The stability of this fixed point also requires that the parameter
$a< (13+\sqrt{17})/12\approx 1.427$
independent of the ratio $\delta/\epsilon$. In spite of the
absence of
renormalization of the forcing correlation, the momentum-shell
approach \cite{Fournier82} yields the same condition.

The system of equations (\ref{beee1}), (\ref{GGGa1}), and
(\ref{G5}) for the fixed points in this multi-charge problem is rather
complicated and thus has several (in general complex-number) solutions, which we do not quote
explicitly here, because they are not physically relevant:
apart from the fixed points listed above there are eight IR unstable real-number fixed points
in the physical region (all $g\ge 0$) of the parameter space, and
several unphysical ones.
Among the unstable fixed points are, in particular, all the
possible candidates to {\em magnetic} fixed points,
i.e. fixed points with a nonvanishing magnetic coupling constant. Therefore,
the conclusion made in Refs.~\cite{Fournier82,Adzhemyan85}
(although on inconsistent grounds) that the RG does
not predict any magnetically driven scaling regime at and near two dimensions,
is confirmed in the double-expansion approach.

It should be noted, however, that the
vanishing of the function $\gamma_5$ renders the linear
combinations of $\gamma$ functions in the functions $\beta_{gb1}$
and  $\beta_{gb2}$ equal: they both contain only
$\gamma_1+2\gamma_2-\gamma_3$ [see Eq (\ref{beee1})]. This has the important
consequence that there is no fixed point with both $g_{b1}$ and $g_{b2}$
nonvanishing: at least one of them must be zero. This, of course,
severely reduces the set of possible magnetic fixed points at the
outset.

 It would be interesting, however, to follow the crossover from this regime
 to the scaling regime governed by the competition of the stable kinetic and magnetic fixed
 points, which exists in three dimensions.
 In the double-expansion approach the space dimension is assumed
 to be close to two, therefore the results obtained above are not
 applicable to this end.
 In the usual $\epsilon$
 expansion the leading-order value of the borderline
 dimension between the two regimes $d_c=(3+\sqrt{649})/10\approx 2.848$~\cite{Fournier82}.

 All the renormalization constants and the RG functions quoted above may be calculated
 also at finite $\delta$.
 The resulting system of fixed-point
 equations allows for a solution in a form of an
 $\epsilon$ expansion (with finite $\delta$) and yields the same
 result as the usual $\epsilon$ expansion at the leading order.
 However, this approach is not self-consistent in the sense that
 the field theory is not renormalizable at finite $\delta$, but
 only in the form of a simultaneous expansion in the coupling
 constants and $\delta$~\cite{Justin89}. Therefore, in order to
 construct an interpolation procedure between the $\epsilon$,
 $\delta$ expansion and the usual $\epsilon$ expansion something
 has to be done with the UV divergences at $d>2$ introduced by the local
 terms in the force correlation functions. This issue will be
 dealt with in Sec. V.

\section{Kolmogorov constant for stochastic magnetohydrodynamics near two dimensions}

The energy spectrum $E(k)$ in magnetohydrodynamics is given by the sum of
equal-time pair correlation functions of the velocity field and
magnetic induction:
\begin{equation}
\label{energyspectrum}
W^{vv}_{nn}(t,{\bf x};t,{\bf x})+W^{bb}_{nn}(t,{\bf x};t,{\bf x})
=2\int_0^\infty\!{\rm d}k\, E(k)\,.
\end{equation}
Note that the energy spectrum is defined through unrenormalized
correlation functions. From (\ref{Rsolution}) and (\ref{energyspectrum}) we infer an
expression for the energy spectrum in terms of the scaling
functions as
\begin{eqnarray*}
E(k)={(d-1)k^{1-4\epsilon/3}\over(4\pi)^{d/2}\Gamma(d/2)}
\left(\frac{g_{v10}u_0\nu_0^3}{{\overline g}_{v1}\overline{u}}\right)^{2/3}\\
\times\left[R_v(0,1; \overline{g})+
Z_3(g)e^{\int_1^s\!{\rm d}x\,\gamma_{3}({\overline g}(x))/x}
R_b(0,1; \overline{g})\right]\,,
\end{eqnarray*}
where
for the running coefficient of viscosity  $\overline{\nu}$ the expression
\begin{equation}
\label{scalingfactor}
\overline{\nu}
=\left(\frac{g_{v10}u_0\nu_0^3}{{\overline g}_{v1}\overline{u}}\right)^{1/3}k^{-2\epsilon/3}
\end{equation}
has been used. The relation (\ref{scalingfactor}) is
a consequence of the connection between the functions
$\beta_{v1}$, $\beta_u$, $\gamma_1$ and $\gamma_2$~\cite{JETP}.

At the kinetic fixed point the spectrum has the form
\begin{eqnarray}
\label{kineticspectrum}
E(k)=\left(\frac{g_{v10}u_0\nu_0^3}{{
g}_{v1}^*u^*}\right)^{2/3}{(d-1)k^{1-4\epsilon/3}\over(4\pi)^{d/2}\Gamma(d/2)}\nonumber\\
\times\left[R_v(0,1; {g}^*)+
Z_3(g)s^{\gamma_3(g^*)}R_b(0,1;{g}^*)\right]\,.
\end{eqnarray}
However, at leading order $R_b(0,1;g)={1\over 2}(g_{b1}+g_{b2})$, and since
at the kinetic fixed point $g_{b1}^*=g_{b2}^*=0$ and $\gamma_3(g^*)>0$, only the scaling
function $R_v$ survives here.

The kinetic and magnetic energy injection rates $\varepsilon_v$ and $\varepsilon_b$
may be expressed as
\begin{eqnarray}
\label{epshi}
\varepsilon_v=\frac{1}{2} \int \frac{\mbox{d}^d {\bf k}}{(2\pi)^d}\,
\langle\,{\bf f}^v({\bf k})\cdot{\bf f}^v(-{\bf k})\,\rangle\,,\nonumber\\
\varepsilon_b=\frac{1}{2} \int \frac{\mbox{d}^d {\bf k}}{(2\pi)^d}\,
\langle\,{\bf f}^b({\bf k})\cdot{\bf f}^b(-{\bf k})\,\rangle\,.
\end{eqnarray}
For our choice of correlation functions after the introduction of simple sharp
cutoffs,
Eq. (\ref{epshi}) yields the relation between
the unrenormalized values of the coupling
constants and the energy injection rates in the form
\begin{eqnarray}
\label{pumping}
 \varepsilon_v=\frac{(d-1)u_0\nu_0^3}{2}\,
 \int_{k_I <k<k_d}\frac{\mbox{d}^d{\bf k}}{(2\pi)^d}\,\nonumber\\
 \times
 \left(\,
 g_{v10}\,k^{2-2\delta-2\epsilon}+g_{v20}\,k^2
 \,\right)\,,\nonumber
 \\
 \varepsilon_b=\frac{(d-1)u_0^2\nu_0^3}{2}\,
 \int_{k_I <k<k_d'}\frac{\mbox{d}^d{\bf k}}{(2\pi)^d}\,\\
 \times
 \left(\,
 g_{b10}\,k^{2-2\delta-2a\epsilon}+g_{b20}\,k^2
 \,\right)\,,\nonumber
\end{eqnarray}
where $k_I$ is the wave number corresponding to the integral
scale, and $k_d$, $k_d'$ are the
characteristic wave numbers of viscous and
resistive dissipation, respectively.

In the stationary state modeling developed isotropic
turbulence, the energy injection
is assumed to take place at large scales.
Therefore, we put the parameters
$g_{v20}$ and $g_{b20}$, which correspond to
small-scale injection of energy,
equal to zero. It should be borne in mind that
the corresponding running coupling constants
are created in the course of renormalization regardless
of the unrenormalized values of these
parameters.
The present perturbative calculation yields only the leading
order in the $\epsilon$, $\delta$ expansion
of the amplitude coefficients in the scaling form of the correlation functions.
Therefore the
coupling constants should be solved from (\ref{pumping})
as functions of $\varepsilon_v$ and
$\varepsilon_b$ also only at leading order of $\epsilon$, $\delta$ expansion.
Thus, we arrive at the relations
\begin{eqnarray}
\label{pumpingconnection}
 \varepsilon_v=
\frac{u_0\nu_0^3\,g_{v10}}{16\pi}\, k_d^{4-2\epsilon}
 \, \,,\nonumber\\
 \varepsilon_b=
\frac{u_0^2\nu_0^3\,g_{b10}}{16\pi}\, (k_d')^{4-2a\epsilon}
\end{eqnarray}
valid for large Reynolds and magnetic Reynolds numbers,
when $k_d/k_I\sim {\rm Re}^{3/4}\gg 1$ and
$k_d'/k_I\sim {\rm Rm}^{3/4}\gg 1$.

Substituting the relations (\ref{pumpingconnection}) in Eq.
(\ref{kineticspectrum}) we arrive at the spectrum in the form
\begin{eqnarray*}
E(k)={(u*)^{1/3}(g_{v1}^*+g_{v2}^*)\over
(2\pi)^{1/3}(g_{v1}^*)^{2/3}}\,{\varepsilon}^{2/3}_v
k^{1-4\epsilon/3}k_d^{4(\epsilon-2)/3}\\
=C_k\,\,
{\varepsilon}^{2/3}_v
\,k^{1- 4 \epsilon/3}
\, k_d^{4(\epsilon - 2)/3}
\end{eqnarray*}
at the leading order of the $\delta$, $\epsilon$ expansion. The
value of the Kolmogorov constant $C_k$ inferred from here
\[
C_k =\frac{2\,\cdot 12^{1/3} \,\epsilon^{1/3}\,
(\epsilon + \delta)^{2/3} }{
(2\epsilon + 3 \delta)^{2/3}}
\]
coincides with that obtained in the case of turbulent advection of
a passive scalar~\cite{Hnatich99}, which is, of course, not
surprising, since the magnetic field is passively advected by the
velocity field in the scaling regime governed by the kinetic fixed
point.

\section{Renormalization with maximum divergences above two dimensions}

We want to maintain the model UV finite for $2\delta=d-2>0$ and simultaneously
keep track of the effect of the additional divergences near two dimensions. To
this end we introduce an additional UV cutoff in all propagators, i.e.
instead of the set (\ref{PropY}) we use the propagators
 \begin{eqnarray*}
 \Delta^{v v'\Lambda}_{mn}({{\bf k}},t)&=&
\theta(t)\,\theta(\Lambda-k)\,P_{mn}({{\bf k}})\,e^{-\nu_0\,k^2 t}\,,
 \\
 \Delta^{bb'\Lambda}_{mn}({{\bf k}},t)&=&
\theta(t)\,\theta(\Lambda-k)\,P_{mn}({{\bf k}})\,
e^{- u_0 \nu_0 k^2 t }\,,
 \\
 \Delta^{vv\Lambda}_{mn}({{\bf k}},t)
 &=&\frac{1}{2}\,\theta(\Lambda-k)
 \,u_0\,\nu_0^2\,P_{mn}({{\bf k}})\,
 e^{ -\nu_0 k^2 | t | }\\
 &\times& \left(\,
 g_{v10}\,k^{-2\epsilon-2\delta}+g_{v20}
 \,\right)\,,\\
 \Delta^{bb\Lambda}_{mn}({{\bf k}},t)&=&
 \frac{1}{2}\,\theta(\Lambda-k)
 \,u_0\,\nu_0^2\,P_{mn}({{\bf k}})\,
 e^{ -u_0 \nu_0 k^2 | t | }\\
 &\times& \left(\,
 g_{b10}\,k^{-2\,a \epsilon- 2\delta} + g_{b20}
 \,\right)\,,
 \end{eqnarray*}
where $\Lambda$ is the cutoff wave number. This change obviously does not affect
the large-scale properties of the model. We would like to
emphasize that the additional cutoff must be introduced uniformly
in all lines in order to maintain the model multiplicatively
renormalizable. An attempt to introduce the cutoff, say, in the
local part of the correlation functions only by the substitution
$k^2\to \theta(\Lambda-k)k^2$ would fail to renormalize the model
multiplicatively, because loop contributions to the complete (dressed)
correlation function would not reproduce such a structure in the wave-vector space.

In contrast with particle field theories we will keep the cutoff
parameter $\Lambda$ fixed, although large compared with the
physically relevant wave-number scale. This introduces an explicit
dependence on $\Lambda$ in the coefficient functions of the RG,
which has to be analyzed separately in the large-scale limit in
the coordinate space. The setup is thus very similar to that of
Polchinski~\cite{Polchinski84}.

The RG equations maintain the previous form (\ref{CS}),
but the coefficient functions become, in general, functions of
the parameters $\mu$ and $\Lambda$ through the dimensionless ratio
$\mu/\Lambda$. Solving the RG equation
by the method of characteristics we obtain the solution
\begin{eqnarray*}
W_{R\ mn}^{vv}(t,{\bf k},\Lambda;g) &=&
P_{mn}({\bf k})\,\overline{ \nu}^2k^{-2\delta}R_v\left(tk^2\overline{\nu},1,{\mu s\over \Lambda};
\overline{g}\right)\,,\\
W_{R\ mn}^{bb}(t,s{\bf k},\Lambda;g) &=&
e^{\int_1^s\!{\rm d}x\,\gamma_{3}(x)/x}\\
&\times&
P_{mn}({\bf k})\, \overline{\nu}^2k^{-2\delta}R_b\left(tk^2\overline{\nu}
,1,{ \mu s\over \Lambda};
\overline{g}\right)\,,
\end{eqnarray*}
where  $\overline{g}$ is now the solution of the Gell-Mann-Low equations:
\[
\frac{d\overline{g}(s)}{d\ln s}= \beta_g
\left[ \overline{g}(s),{\mu s\over \Lambda}\right]\,,
\]
with the $\beta$ functions explicitly depending on $s$, the dimensionless wave number.

Above two dimensions an UV renormalization of the model would
require and infinite number of counter terms and in this sense it
is not renormalizable in the limit $\Lambda\to \infty$.
To avoid dealing with these formal complications, we keep the additional
cutoff $\Lambda$ fixed (although large), and choose the
renormalization procedure according to the principle of maximum
divergences~\cite{Honkonen89}: the same terms of the action are
renormalized as in the two-dimensional case in the previous section
(\ref{kontra}), but the renormalization constants may
have an explicit dependence on the scale-setting
parameter through the ratio $\mu/\Lambda$.
The two counter terms
$\int\!\mbox{d} x\,
\big[\frac{1}{2}\,(1-Z_4)
u {\nu^3} g_{v2}\,{\mu}^{-2\delta}\,
{{\bf v}}'{\nabla^2} {{\bf v}}'
+\frac{1}{2}\,
(1- Z_5 )\,u^2 {\nu^3} g_{b2}\,{\mu}^{-2\delta}\,
{{\bf b}}'{\nabla^2} {{\bf b}}'
\, \big]
$
are superfluous in the sense that in the limit $\mu/\Lambda\to 0$
the contribution to the Green functions of the graphs containing
the coupling constants $g_{v2}$ and  $g_{b2}$ remains finite
provided $2\delta=d-2$ is fixed and finite and the other
counter terms are properly chosen. This is
guaranteed by Polchinski's theorem~\cite{Polchinski84}. We retain
these counter terms in order to have a possibility to pass to the
limit $\delta\to 0$ in the RG equations.

The presence of the explicit cutoff implies some technical difficulties
in the calculation of the renormalization constants in the
traditional field-theoretic approach, which arise because
we are dealing with massless vector fields. It turns out that
the coefficient functions of the
RG equation are simplest in a
renormalization procedure, which is similar to the
momentum-shell renormalization of Wilson~\cite{Wilson74}.
If we were calculating over the whole wave-vector space without an
explicit UV cutoff, there would not be much difference between the
effort required in both approaches. The presence of the UV cutoff
makes calculations with nonvanishing external wave vectors rather
tedious.

Let us remind that the choice of a renormalization procedure
basically is the choice of the rule according to which the counter-term
contributions are extracted from the perturbation expansion
of the Green functions of the model.
The usual field-theoretic prescription goes as
follows~\cite{Collins85}:
Consider a 1PI graph $\gamma$, let $R(\gamma)$ be the renormalised value
of the graph, and let $\overline{R}(\gamma)$ the value of the graph with
subtracted counter terms of all the subgraphs, then
\begin{equation}
\label{R}
R(\gamma)=\overline{R}(\gamma)- T\overline{R}(\gamma)
\end{equation}
where the operator $T$ extracts the counter-term contribution
from $\overline{R}(\gamma)$. Usually $\overline{R}(\gamma)=\gamma$ on
one-loop graphs, and the renormalization scheme is specified by
the action of $\overline{R}(\gamma)$ on multiloop graphs together
with the definition of the operator $T$. The counter terms may
then be constructed recursively with the use of (\ref{R}) and the
definitions of $T$ and $\overline{R}(\gamma)$. There is rather large
freedom in the choice of the counter-term operator, but to arrive
at Green functions finite in the limit $\Lambda\to \infty$
in two dimensions -- which we want to have a connection with the
double expansion -- the operation $\overline{R}$ must be chosen
properly.

Here, we have used a renormalization procedure, in which the
operation $\overline{R}(\gamma)$ is standard~\cite{Collins85}, and the
subtraction operator $T$ is defined as follows:
let $F_\gamma(\omega,{\bf k},\Lambda)$ be the function of external
frequencies and wave-vectors (which also depends on the cutoff parameter $\Lambda$)
corresponding to the expression $\overline{R}(\gamma)$ (this is not a
1PI graph, in general).
The subtraction operator $T$ extracts the same set of terms
of the Maclaurin-expansion in the external wave-vectors,
which generate the counter terms (\ref{kontra}), from the {\em
difference} $F_\gamma(\omega,{\bf k},\Lambda)-F_\gamma(\omega,{\bf
k},\mu)$. These coefficients of the Maclaurin-expansion are
calculated at vanishing external frequencies and wave-vectors.
It should be noted that the coefficients of this Maclaurin
expansion of the function $F_\gamma(\omega,{\bf k},\Lambda)$
itself may not exist in the limit $\omega\to 0$ in this "massless" model, but the difference
$F_\gamma(\omega,{\bf k},\Lambda)-F_\gamma(\omega,{\bf
k},\mu)$ allows for a Maclaurin expansion finite in the limit $\omega\to 0$ to the order required
for the renormalization. The counter-term operator $T$ and the combinatorics of the
renormalization procedure for higher-order graphs may then be
constructed in the standard fashion. Although this is actually not needed
in the present one-loop calculation, the very possibility of this
extension is necessary to guarantee that the renormalization renders the model finite
in the limit $\Lambda\to\infty$ in two dimensions.

Effectively, at one-loop order this prescription reduces the region of
integration to the momentum shell $\mu<k<\Lambda$, which leads to
the same calculation as in the momentum-shell renormalization. In
higher orders, however, our renormalization scheme does not coincide with
the momentum-shell renormalization.
The point of the present renormalization procedure is that without some sort of IR cutoff the
subtraction at zero momenta and frequencies is, in general, not
possible in a massless model, whereas a subtraction at vanishing
frequencies and external momenta of the order of $\mu$ leads to
much more complicated calculations due the heavy index structure.

At one-loop accuracy in this scheme the $\gamma$ functions are
\begin{eqnarray}
\gamma_1&=&\frac{1}{2B}\Bigl[\left(d^2-d-2\epsilon\right)u\,g_{v1}+\left(d^2+d-4+2
a \epsilon\right) g_{b1}\nonumber\\
&+&
 \left(d^2-2\right)\left(u\,g_{v2} + g_{b2}\right)\Bigr]\,,
\nonumber\\
\gamma_2&=&\frac{1}{(1+u)B}\Bigl[\left(d-1\right)\left(d+2\right)\left(g_{v1}+
g_{v2}\right)\nonumber\\
&+&\left(d+2\right)\left(d-3\right)\left(g_{b1}+g_{b2}\right)\Bigr]\,,
 \nonumber\\
\gamma_3&=&{2\over B}\left[g_{b1}+g_{b2}-g_{v1}-g_{v2}
\right]\,, \label{GGGa2}\\
 \gamma_4&=&\frac{d^2-2}{2 g_{v2}B}\left[u\left(g_{v1}+g_{v2}\right)^2+
 \left(g_{b1}+g_{b2}\right)^2\right]\,,
 \nonumber\\
\gamma_5&=&\frac{2(d-2)(d+2)}{g_{b2}(1+u)B}
\left(g_{b1}+g_{b2}\right)\left(g_{v1}+g_{v2}\right)\,,\nonumber
\end{eqnarray}
where $B=d(d+2)\Gamma(d/2)(4\pi)^{d/2}$. These expressions reveal
an additional advantage of this renormalization scheme: at
one-loop order there is no explicit dependence on $\mu/\Lambda$ in
the coefficient functions of the RG. At one-loop level
a direct comparison with the expressions obtained in the
framework of the Wilson RG is also possible:
the dependence on $g_{v1}$,  $g_{b1}$, and $u$ of the $\beta$ functions
$\beta_{gv1}$, $\beta_{gb1}$ and $\beta_{u}$ corresponding to
(\ref{GGGa2}) coincides with that of their
counterparts of Ref.~\cite{Fournier82} up to notation.

The set of $\beta$ functions generated by (\ref{GGGa2}) allows for
a fixed-point solution in the form of an $\epsilon$ expansion.
Little reflection shows that the fixed-point equations in this
case have a self-consistent solution with the following
leading-order behavior: $u=O(1)$,
$g_{v1}=O(\epsilon)$, $g_{b1}=O(\epsilon)$,
$g_{v2}=O(\epsilon^2)$ and $g_{b2}=O(\epsilon^2)$. The
actual fixed-point values of $g_{v1}$, $g_{b1}$ and $u$ in the
$\epsilon$ expansion as well as
the stability regions with respect to $\epsilon$ are determined by
the same set of equations as in the earlier momentum-shell~\cite{Fournier82}
and field-theoretic~\cite{Adzhemyan85} calculation
above two dimensions. The stability condition with respect to the
dimension of space of these fixed points is, as expected, $d>2$.

It should be noted that the function $\gamma_5$ is finite in the
set (\ref{GGGa2}), whereas in the double-expansion approach it was
equal to zero [Eq.  (\ref{G5})]. This means that magnetic fixed
points with both $g_{b1}$ and $g_{b2}$ may exist. In
fact, there is one such fixed point stable at high dimensions of
space which gives rise to a magnetically driven scaling regime.
This fixed point may be found in the $\epsilon$ expansion, and we
have also investigated its stability numerically. Technically
speaking, the appearance of a magnetic fixed point with both
magnetic couplings nonvanishing would be a completely expected thing to
happen in the two-loop approximation, since we have not found any
symmetry reasons or the like to prevent the renormalization of the
magnetic forcing at higher orders. Thus, to investigate this
effect consistently in the $\epsilon$ expansion would
require a full two-loop renormalization of the model, which is
beyond the scope of the present analysis.

On the other hand, in the case $\delta\sim\epsilon\to 0$ the set
of coefficient functions (\ref{GGGa2}) yields the coefficient
functions (\ref{GGGa1}) of the double expansion of the previous
section. Therefore, we think that it is not totally unreasonable to use the
set of coefficient functions (\ref{GGGa2}) for an analysis of the
RG fixed points for all dimensions $d\ge 2$.

\begin{figure}
\begin{center}
\epsfxsize=8truecm
\epsffile{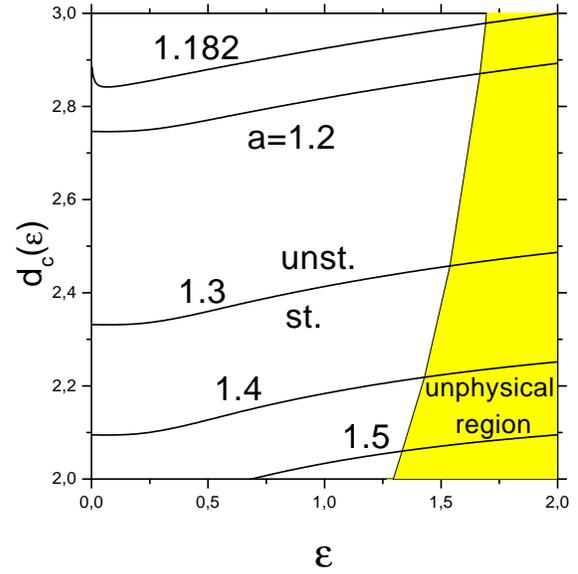}
\end{center}
\caption{The borderline dimension $d_c$ between the stability regions of the kinetic fixed point
of the RG equations (\ref{beee1}) and (\ref{GGGa2}) for
magnetic forcing-decay parameter $a$ near the double-expansion treshold $a=1.427$.
This plot reveals the strong dependence of the borderline
dimension on the parameter $a$. The shaded region on the right
corresponds to values $\epsilon>2/a$, for which the forcing
correlation function in the powerlike form (\ref{silaN}) leads to intractable IR
divergences, and a corresponding IR cutoff (magnetic integral
length scale) must be introduced.}
\label{fig1}
\end{figure}

\begin{figure}
\begin{center}
\epsfxsize=8truecm
\epsffile{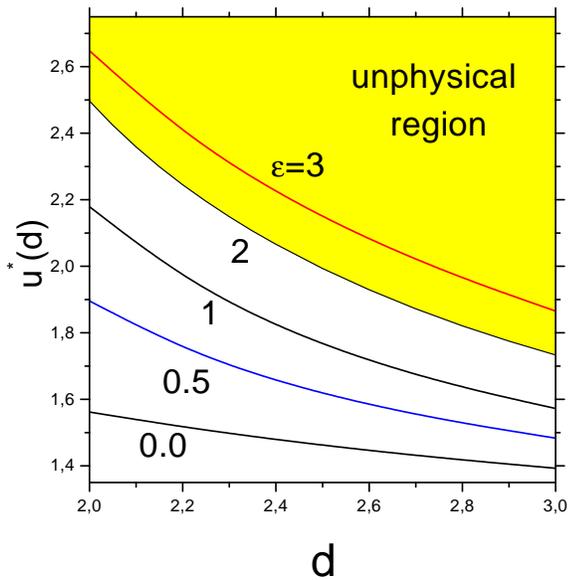}
\end{center}
\caption{The fixed-point value of the inverse magnetic Prandtl number $u^*$ as a
function of the space dimension $d$ and the decay parameter
$\epsilon$. The lowest curve corresponds to the leading order in
the $\epsilon$ expansion, which is not affected by thermal fluctuations.
The shaded region in the upper part
of the plot
corresponds to values $\epsilon>2$, for which an IR cutoff (kinetic integral
length scale) must be introduced in the
correlation function (\ref{silaN}).}
\label{fig2}
\end{figure}

We have investigated the stability of the kinetic fixed point
given by (\ref{beee1}) and (\ref{GGGa2}) numerically for a finite range of
$\epsilon$ with results depicted in
Figs. \ref{fig1} and \ref{fig2}. We think that this calculation,
although it is a somewhat uncontrollable approximation,
exhibits the effect of the thermal (short-range) fluctuations of
the fields qualitatively correctly.
The stability of the kinetic scaling regime is
strongly affected by the behavior of magnetic fluctuations: from
Fig.\ref{fig1} it is seen that the steeper falloff of the correlations of
the magnetic forcing in the wave-vector space compared with that
of the kinetic forcing the lower is the space dimension, above
which the kinetic fixed point is stable. In particular, when the
parameter $a>1.427$ the kinetic fixed point ceases to be stable
even in two dimensions. In three dimensions the kinetic scaling
regime is stable against magnetic forcing, when $a<1.15$.

\begin{figure}
\begin{center}
\epsfxsize=8truecm
\epsffile{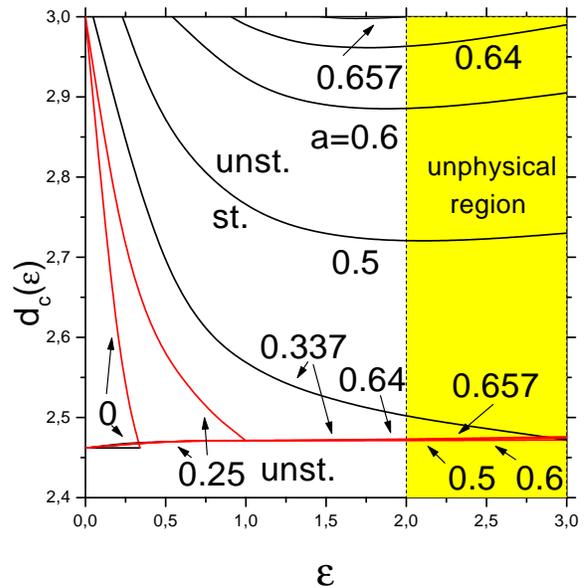}
\end{center}
\caption{The borderline dimension $d_c$ between the stability regions of the magnetic fixed point
of the RG equations (\ref{beee1}) and (\ref{GGGa2}) for
magnetic forcing-decay parameter $a<1$. For sufficiently small values of $a$ the magnetic
fixed point is unstable for any finite value of $\epsilon$, but
the region of stability grows with the growth of $a$ so that for
$a>0.658$ the magnetic point becomes stable even in three
dimensions for finite values of $\epsilon$. The shading shows the
region, where $\epsilon>2$, in which the powerlike correlation
function (\ref{silaN}) cannot be consistently used.}
\label{fig3}
\end{figure}

\begin{figure}
\begin{center}
\epsfxsize=8truecm
\epsffile{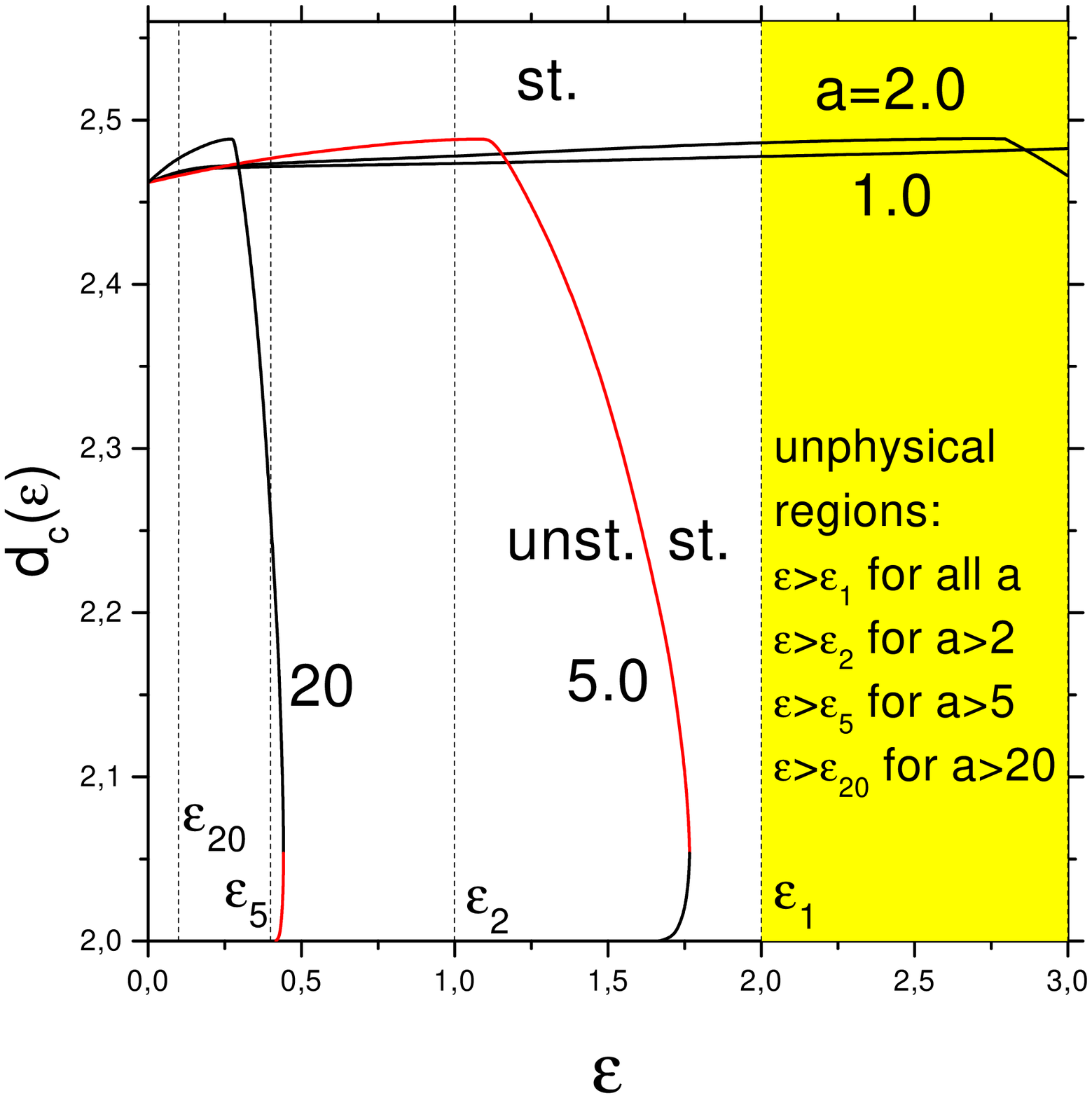}
\end{center}
\caption{The borderline dimension $d_c$ between the
stability regions of the magnetic fixed point of the RG equations
(\ref{beee1}) and (\ref{GGGa2}) for large values of the magnetic
forcing-decay parameter $a>1$. The shaded area and half-planes with vertical
dashed border lines show
regions, where $\epsilon>2/a$, in which the powerlike correlation
function (\ref{silaN}) cannot be consistently used.
%It is a noteworthy feature of this
%plot that the curve corresponding to the equality
%$a\epsilon=2$ is practically a straight line.
}
\label{fig4}
\end{figure}

The monotonic growth of the kinetic-fixed-point value of the
inverse magnetic Prandtl number $u^*$ as a function of the kinetic forcing-decay parameter
in a fixed space dimension
is depicted in Fig. \ref{fig2}. The plot shows also that
$u^*$ is a monotonically decreasing function of the space
dimension at fixed $\epsilon$. The lowest-lying curve corresponds
to the leading order of the $\epsilon$ expansion~\cite{Fournier82}
\[
u^*=\frac{1}{2}\left[ -1+
  \sqrt{1+\frac{8\,(d+2)}{d}}\,\right]\,.
\]
We are particularly interested in the stability of the magnetic
fixed point, and have carried out extensive numerical calculations
of the stability of this fixed point as a function of $\epsilon$
and  the space dimension $d$. The results are plotted in Figs.
\ref{fig3} and  \ref{fig4}.

In Fig. \ref{fig3} the magnetic forcing-decay parameter $a<1$
(i.e. the kinetic-forcing correlations fall off steeper in the
wave-vector space than those the magnetic forcing) and it is seen
that for very small $a$ a slowly enough decaying kinetic forcing
renders the magnetic scaling regime unstable. In particular, this treshold is very
small in three dimensions. With the growth of
$a$ a strip of stability of the magnetic fixed point in the
$\epsilon$, $d$ plane appears such that the magnetic regime
remains stable in three dimensions for all allowed kinetic forcing
patterns.  It is also seen that the magnetic
fixed point is persistently unstable at $d\le 2.46$ for all
$\epsilon$. This borderline dimension should be compared with
that given by the $\epsilon$ expansion $d_c=2.85$. From the
solution it can be seen that this
significant discrepancy is due to the appearance of a stable
magnetic fixed point completely different from that found in the
$\epsilon$ expansion: in the latter the magnetic fixed point
is given by $g_{v1}^*=g_{v2}^*=g_{b2}^*=u^*=0$ and
$g_{b1}^*=4d(d+2)\Gamma(d/2)(4\pi)^{d/2}a\epsilon/(d^2-3d-32)$,
whereas at the magnetic fixed point, whose stability is plotted in
Figs. \ref{fig3} and \ref{fig4}, only $g_{v1}^*=u^*=0$ with
nonvanishing fixed-point values of the other couplings. Thus, the
lowering of the borderline dimension of stability of the magnetic
scaling regime is a
strong effect of the thermal fluctuations described by the
short-range terms in the forcing correlation functions.  Fig. \ref{fig4}
shows the lower boundary of the stability region of the magnetic
fixed point for large values of $a$, when magnetic-forcing
correlations fall off much faster that kinetic-forcing
correlations in the wave-vector space. A remarkable feature of
both plots is the insensitivity of the lower border of the stability strip to the
value of magnetic forcing-decay parameter $a$.

\section{Conclusions}

In conclusion, we have carried out a RG
analysis of the large-scale asymptotic behavior of the solution
of stochastically forced magnetohydrodynamic equations for all
space dimensions $d\ge 2$. We have taken into account the
additional divergences appearing in two dimensions ignored or
improperly treated in previous work. In a two-parameter expansion
scheme we have found three infrared-stable fixed points in the physically relevant
region of the parameter space spanned by the forcing parameters
and the inverse magnetic Prandtl number. Anomalous scaling behavior is
brought about in the basins of attraction of the thermal and
kinetic fixed points, the former of which is due to thermal
fluctuations, and the latter due to long-range correlated random
forcing of the Navier-Stokes equation. The thermal fixed
point is related to the anomalous asymptotic behavior due to thermal fluctuations
near two dimensions. With a proper choice of the
force-correlation function the regime governed by the kinetic fixed point
may be related to developed
isotropic turbulence in a conducting fluid. We have obtained the stability
condition
of the kinetic fixed point near two dimensions with respect to the magnetic
forcing-decay parameter as $a<1.427$,
which coincides with that of Ref. \cite{Fournier82}, but differs
from the result of Ref. \cite{Adzhemyan85}.

We have also put forward an interpolation scheme, which reproduces the
earlier results in the case of an $\epsilon$
expansion in any fixed space dimension $d>2$, and the results of the present paper
in the two-parameter
expansion in $\epsilon$ and $2\delta=d-2$ near two dimensions.
Using this interpolation approach we have qualitatively analysed the dependence
of the stability of the kinetic and magnetic
scaling regimes on the forcing-decay parameters and the dimension of
space and found that thermal fluctuations drastically lower the borderline
dimension of stability of the magnetic scaling regime.

\acknowledgements

M. H. gratefully acknowledges the hospitality of the N. N. Bogoliubov
Laboratory of Theoretical Physics at JINR Dubna and the Department of Physics
of the University of Helsinki, Finland. J. H. thanks the Institute for Experimental
Physics of the Slovak Academy of Sciences in Ko\v{s}ice for
hospitality.

This work was supported in part by Slovak Academy of Sciences (Grant No. 7232) and
by the Academy of Finland (Grant No. 64935).

% ###  Apendixy:  ###

\begin{references}

\bibitem{Frisch95}
U. Frisch, {\em Turbulence: the legacy of A. N. Kolmogorov}
(Cambridge University Press, Cambridge, 1995).

\bibitem{Adzhemyan99}
L. Ts. Adzhemyan, N. V. Antonov, and A. N. Vasiliev, {\em
The Field Theoretic Renormalization Group in Fully Developed
Turbulence} (Gordon and Breach, Amsterdam, 1999).



\bibitem{Fournier82}
J. D. Fournier, P. L. Sulem, and A. Pouquet,
J. Phys. A {\bf 15}, 1393  (1982).
%Fournier J.D., Sulem P.L., and Poquet A.:
%J. Phys. A: Math. Gen. {\it 15} (1982) 1393.

\bibitem{Adzhemyan85}
L. Ts. Adzhemyan, A. N. Vasil'ev, and M. Hnatich,
Teor.  Mat. Fiz. {\bf 64}, 196 (1985).
%Adzhemyan L.Ts., Vasilyev A.N., and  Hnatich M.:
%Teor. i Matemat. Fiz. {\it 64} (1985) 196.

\bibitem{Honkonen96}
J. Honkonen and M. Yu.  Nalimov, Z. Phys. B
{\bf 99}, 297 (1996).
%  Honkonen J., Nalimov M. Yu.: Z. Phys. B
%  {\it 99} (1996) 297.

\bibitem{Liang93}
W. Z. Liang and P. H. Diamond, Phys. Fluids B {\bf 5}, 63 (1993).

\bibitem{Kim99}
C.-B. Kim and T.-J. Yang, Phys. Plasmas {\bf 6}, 2714 (1999).



\bibitem{Forster77}
D. Forster, D. R. Nelson, and M. J. Stephen,
Phys. Rev. A {\bf 16} 732 (1977).


\bibitem{Wilson74}
K. G. Wilson and J. Kogut, Phys. Rep. {\bf 12}, 75 (1974).

\bibitem{Camargo92}
S. J. Camargo and H. Tasso, Phys. Fluids B {\bf 4}, 1199 (1992).

\bibitem{Horvath95}
M. Hnatich, D. Horv{\'a}th, R. Seman\v{c}\'{\i}k, and
M.  Stehl\'{\i}k, Czech. J. Phys. {\bf 45}, 91 (1995).
 % Hnatich  M., Horv{\'a}th M., Seman\v{c}\'{\i}k R.,
 % Stehl\'{\i}k M.: Czech. J. Phys. {\it 45} (1995) 91.
 % " On the critical regimes in the theory of randomly
 % forced magnetohydrodynamic turbulence."
 % Vol.45, No.1, 1995


\bibitem{Honkonen98}
J. Honkonen, Phys. Rev. E {\bf 58}, 4532 (1998).

\bibitem{Hnatich99}
M. Hnatich, J. Honkonen, D. Horv{\'a}th, and R. Seman\v{c}\'{\i}k,
Phys. Rev. E {\bf 59}, 4112 (1999).

 \bibitem{Vasilyev83}
 L. Ts. Adzhemyan, A. N. Vasil'ev, and Yu. M. Pis'mak,
 Teor.  Mat. Fiz. {\bf 57}, 268 (1983).
 %Adzhemyan L.Ts, Vasilyev A.N., Pis'mak Yu.M.:
 %Teor. i Matemat. Fiz. {\it 57} (1983) 268.
 % No.2, p.268-281.

 \bibitem{DeDominicis76}
C. De Dominicis, J. Phys. (Paris) {\bf 37}, Suppl C1, 247 (1976).

\bibitem{Janssen76}
H. K. Janssen, Z. Phys. B {\bf 23}, 377 (1976).


\bibitem{Justin89}
 J. Zinn-Justin,
  {\em Quantum Field Theory and Critical Phenomena}
  (Oxford University Press, Oxford, 1989).

\bibitem{JETP}
L. Ts. Adzhemyan, N. V. Antonov, and A. N. Vasil'ev, Zh. Eksp. Teor. Fiz.
{\bf 95}, 1272 (1989) [Sov. Phys. JETP {\bf 68}, 733 (1989)].

\bibitem{Honkonen89}
%Honkonen J, Nalimov M. Yu: Crossover between field theories with
%short-range or long-range exchange or correlations.
J. Honkonen and M. Yu. Nalimov,
J. Phys. A {\bf 22},  751 (1989).



\bibitem{Polchinski84}
J. Polchinski, Nucl. Phys. B {\bf 231}, 269 (1984).



\bibitem{Collins85}
J. Collins, {\em Renormalization} (Cambridge University Press, Cambridge, 1985)


\end{references}
\end{document}